# Hyperbolic Phonon-Plasmon Modes in Grounded Graphene-hBN Heterostructures for Mid-Infrared Applications

**Mohammad Bagher Heydari [1,*], Majid Karimipour [2], Morteza Mohammadi Shirkolaei [3]**

[1,*] School of Electrical Engineering, Iran University of Science and Technology (IUST), Tehran, Iran
[2] Department of Electrical Engineering, Arak University of Technology, Arak, Iran
[3] Department of Electrical Engineering, Shahid Sattari Aeronautical University of Science and Technology, Tehran, Iran

[*]Corresponding author: mo_heydari@alumni.iust.ac.ir

**Abstract:** In recent years, the hybridization of hyperbolic van der Waals heterostructures with plasmonic two-dimensional nano-materials is one of the interesting research areas at THz frequencies due to the coupled features of the hybrid structure. This article investigates the propagation of tunable surface phonon-plasmon polaritons ($SP^3$) in grounded hybrid graphene-hexagonal Boron Nitride (hBN) heterostructures. An analytical model is presented for the proposed structure, by applying the boundary conditions for the electromagnetic components in the various regions, to derive an exact dispersion relation. The structure is simulated and the results are reported in the upper Reststrahlen band. A good agreement is seen between simulation and analytical results, which shows the high accuracy of our mathematical relations. The tunability of the heterostructures is shown by studying the effect of chemical potential on the performance of the structure. A high value of the figure of merit, i.e. FOM=190, is reported at the frequency of 48.3 THz. The presented study can open the way for the design of novel THz devices for future nano-plasmonic applications.

## 1. Introduction

Nowadays, van der Waals heterostructures are one of the leading topics in physics and material science since they open a huge potential for to design and fabrication of innovative devices for next-generation plasmonics [1]. Among these 2D materials, graphene has attracted immense interest since 2004 due to its fascinating features in a new, emerging research area called "Graphene Plasmonics" [2]. The ability to support highly-confined plasmonic modes at smaller scales, low-loss light-matter interactions, and presenting tunability plasmons via electrostatic or magnetostatic bias, will pave the way to introduce and design new applications such as couplers [3-5], filters [6-8], resonators [9-11], circulators [12-15], waveguides [16-25], sensing [26-31], and imaging [32, 33]. Graphene-based waveguides have various structures such as planar [17, 25, 34-64], cylindrical [45, 65-70], and elliptical structures [24, 71-73].

Hybridization of graphene with other Van der Waals structures can open new insight to the researchers to effectively control the properties of propagating waves inside the hybrid structure and also will give the designer a high degree of freedom. Hexagonal Boron Nitride (hBN) is a polar dielectric, with exotic features such as high thermal conductivity, which can be easily combined with other 2D materials to give new hybrid properties [74-76]. This material can confine phonon polaritons in a very small area, compared to the wavelength, due to its negative permittivity in the mid-infrared region. Therefore, the hybridization of graphene with the hBN medium can produce new modes called "Coupled Plasmon-Phonon modes" [77-80]. These hybrid surface plasmon-phonon polaritons ($HSP^3$) have potential applications in the THz region such as plasmon-induced transparency [81], negative refractive [82], planar focusing [83], and topological transition [84]. This combination will also provide a clean substrate for higher carrier mobility of plasmonic components. In the literature, the first study on the propagating $HSP^3$ in basic platforms is done by [75]. The investigation of $HSP^3$ excitation in the nano-resonators is performed by [77]. A novel type of $HSP^3$ based on acoustic waves arising from a piezoelectric material is introduced in [85]. The papers [86, 87] have investigated THz metamaterials containing hybrid graphene-hBN heterostructures.

In this paper, a new graphene-hBN heterostructure is introduced and investigated in order to obtain higher values of FOM and effectively control the features of propagating plasmon-phonon modes via chemical potential and hBN dielectric characteristics. Then, we present a new mathematical model for the proposed structure by starting from Maxwell's equations and applying boundary conditions. To the best of our knowledge, no published work is reported for the study of $SP^3$ in grounded graphene-hBN heterostructures. It supports coupled plasmon-phonon modes in the mid-infrared region. It should be noted that a grounded heterostructure is investigated here because we intend to confine the coupled plasmon-phonon modes at $z > 0$. As the electric field is zero inside PEC, thus the propagating modes cannot penetrate inside it.

The remainder of the paper is organized as follows. In section 2, the studied structure will be introduced and its analytical model will be presented. An exact dispersion relation is derived in this section. Then, in section 3, the numerical results are reported and investigated in detail. We will compare the simulation and analytical results to check the validity of the proposed model in this section. A high value of FOM=190 is reported at the frequency of 48.3 THz for the studied device. Finally, section 4 concludes the article.

## 2. The Proposed Structure and its Analytical Model

Fig. 1 illustrates the schematic of the proposed structure, where a graphene sheet has been located on grounded $hBN\text{-}SiO_2\text{-}Si$ layers. The graphene conductivity can be modeled by the following relation [88]:

$$\sigma\left(\omega,\mu_g,\Gamma,T\right)=\frac{-je^2}{4\pi\hbar}Ln\left[\frac{2\left|\mu_g\right|-(\omega-j2\Gamma)\hbar}{2\left|\mu_g\right|+(\omega-j2\Gamma)\hbar}\right]+\frac{-je^2K_BT}{\pi\hbar^2(\omega-j2\Gamma)}\left[\frac{\mu_g}{K_BT}+2Ln\left(1+e^{-\mu_g/K_BT}\right)\right] \qquad (1)$$

In (1), $\Gamma$ is the scattering rate, $T$ is the temperature, and $\mu_g$ is the chemical potential of graphene. Furthermore, $\hbar$ is the reduced Planck's constant, $K_B$ is Boltzmann's constant, $\omega$ is radian frequency, and $e$ is the electron charge [88].

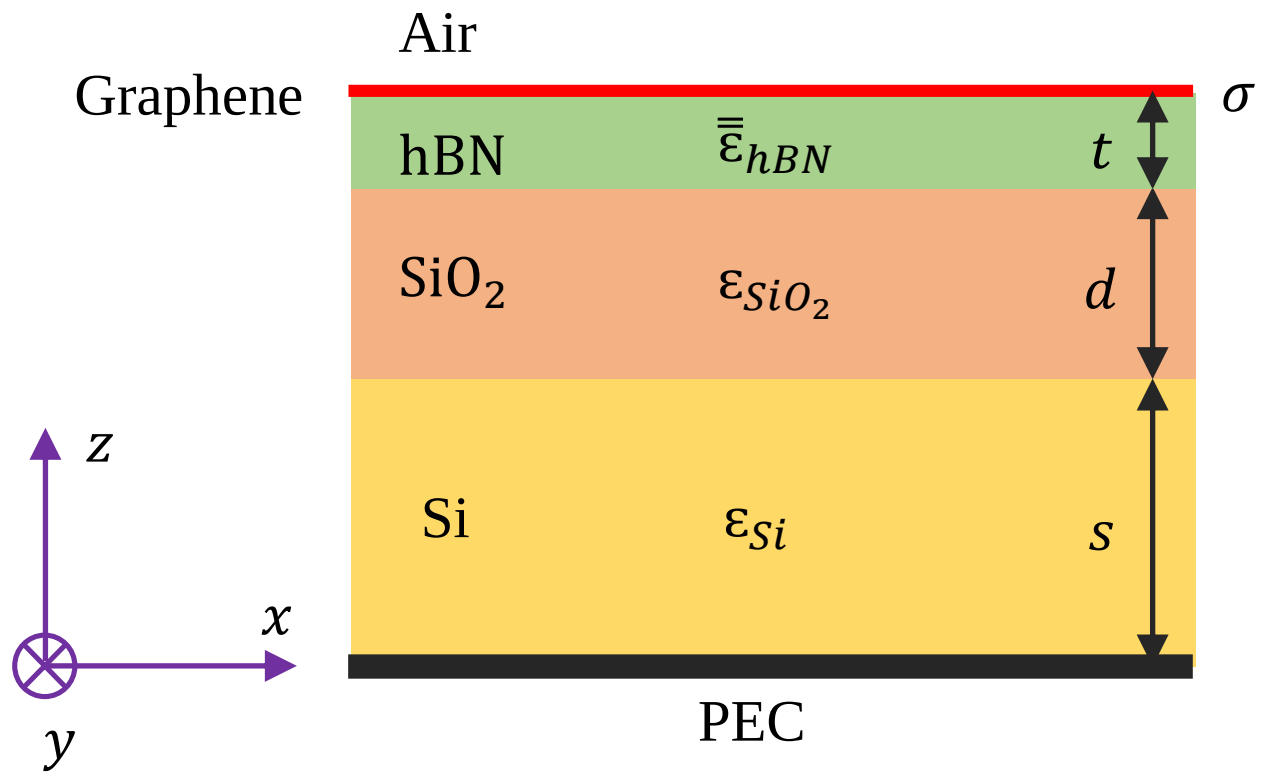

**Fig. 1.** The schematic of the proposed structure.

hBN is a polar dielectric, in which its permittivity can be expressed as follows [75]:

$$\varepsilon_m(\omega) = \varepsilon_{\infty,m} + \varepsilon_{\infty,m} \cdot \frac{(\omega_{LO,m})^2 - (\omega_{TO,m})^2}{(\omega_{TO,m})^2 - \omega^2 - j\omega\Gamma_m} \tag{2}$$

In relation (2), $m = \parallel \; or \; \perp$ is related to the transverse and z-axis, respectively. Moreover, $\omega_{LO}, \omega_{TO}$ show the LO and TO phonon frequencies, respectively, in which each frequency has two values in the upper and lower Reststrahlen bands: $\omega_{LO,\perp} = 24.9\, THz, \omega_{TO,\perp} = 23.4\, THz, \omega_{LO,\parallel} = 48.3\, THz,\ \omega_{TO,\parallel} = 41.1\, THz$. Also, $\Gamma_m$ is a damping factor ($\Gamma_\perp = 0.15\, THz, \Gamma_\parallel = 0.12\, THz$) and $\varepsilon_m$ is related to the high-frequency permittivity ($\varepsilon_{\infty,\perp} = 4.87, \varepsilon_{\infty,\parallel} = 2.95$) [75]. Fig. 2 shows the real part of hBN permittivity. As seen in this figure, two modes exist that are related to type-I and type-II hyperbolicity of hBN: the upper Reststrahlen and the lower Reststrahlen band. In fig. 2, these frequency bands are shown with brown and green rectangles, respectively.

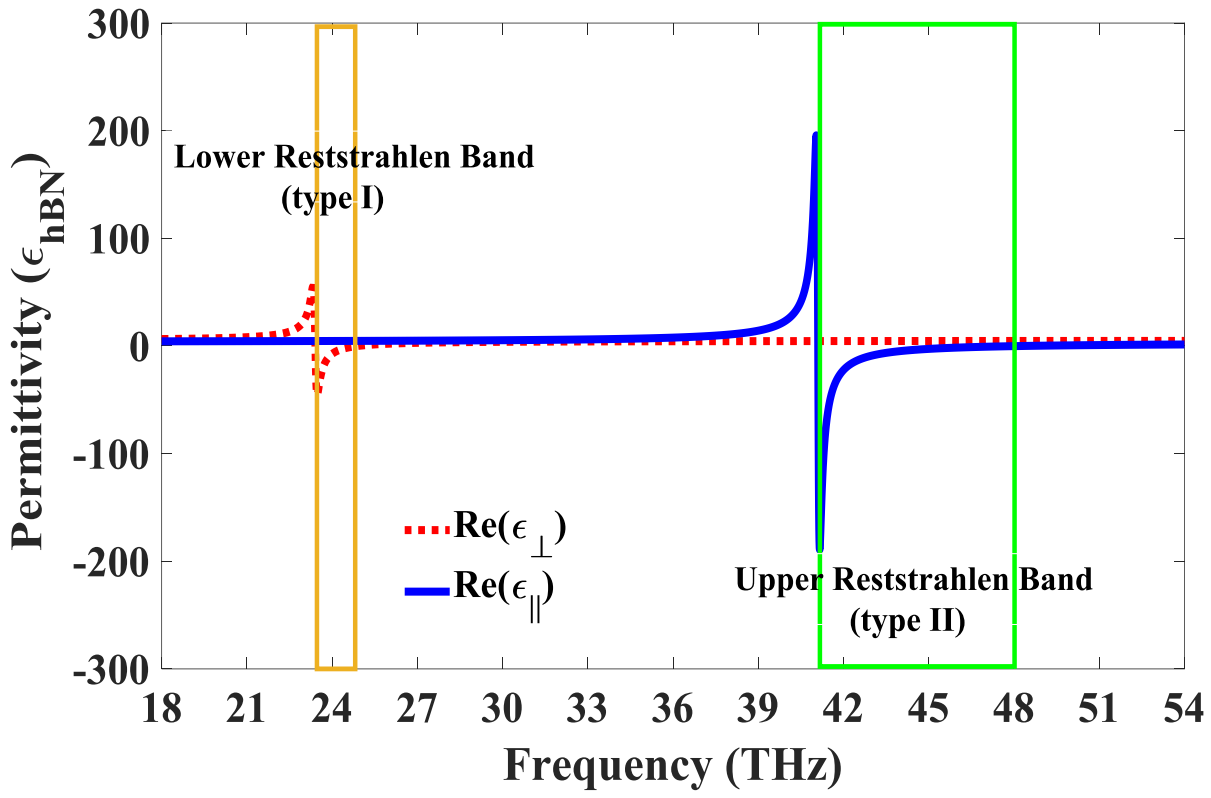

**Fig. 2.** The permittivity of hBN versus frequency. The lower and upper Reststrahlen bands are shown in this figure.

To analyze the structure, we consider TM mode ($E_x, E_z, H_y$) propagating in the x-direction ($e^{j\beta x}$). Maxwell's equations inside the hBN layer are written as follows (suppose$e^{-j\omega t}$):

$$\nabla \times \boldsymbol{E} = j\omega\mu_0 \boldsymbol{H} \tag{3}$$

$$\nabla \times \boldsymbol{H} = -j\omega\varepsilon_0 \overset{=}{\varepsilon} . \boldsymbol{E} \tag{4}$$

By doing some mathematical calculations, the wave equation inside the hBN is obtained:

$$\frac{d^2 H_y}{dz^2} - \gamma_{hBN}^2 H_y = 0 \tag{5}$$

Where the propagation constant is defined as:

$$\gamma_{hBN}^2 = \beta^2 \frac{\varepsilon_\perp}{\varepsilon_\parallel} - k_0^2 \varepsilon_\perp \tag{6}$$

To obtain the transverse component of electric fields, i.e. $E_x, E_z$, we utilize Maxwell's equations:

$$E_x = \frac{1}{j\omega\varepsilon_0\varepsilon_\perp} \frac{\partial H_y}{\partial z} \tag{7}$$

$$E_z = \frac{-\beta}{\omega\varepsilon_0\varepsilon_\parallel} H_y \tag{8}$$

Now, the electromagnetic relations are completed for the hBN layer. It is worth to be mentioned that electromagnetic components inside the dielectric media (Si and $SiO_2$ layers) can be calculated by using the above relations for $\varepsilon_\perp = \varepsilon_\parallel = \varepsilon_{SiO_2}\ or\ \ \varepsilon_\perp = \varepsilon_\parallel = \varepsilon_{Si}$. Let us write the magnetic component ($H_y$) of propagating waves in various regions:

$$H_y(z)=\begin{cases} Ae^{-j\gamma_{Si}z}+Be^{+j\gamma_{Si}z} & 0<z<s \\ Ce^{-j\gamma_{SiO_2}z}+De^{+j\gamma_{SiO_2}z} & s<z<s+d \\ Ee^{-j\gamma_{hBN}z}+Fe^{+j\gamma_{hBN}z} & s+d<z<s+d+t \\ Ge^{-j\gamma_0 z} & z>s+d+t \end{cases} \tag{9}$$

Where $E_x$ can be obtained from (7):

$$E_x(z)=\frac{-1}{\omega\varepsilon_0}\begin{cases} \frac{\gamma_{Si}}{\varepsilon_{Si}}\left(Ae^{-j\gamma_{Si}z}-Be^{+j\gamma_{Si}z}\right) & 0<z<s \\ \frac{\gamma_{SiO_2}}{\varepsilon_{SiO_2}}\left(Ce^{-j\gamma_{SiO_2}z}-De^{+j\gamma_{SiO_2}z}\right) & s<z<s+d \\ \frac{\gamma_{hBN}}{\varepsilon_{hBN}}\left(Ee^{-j\gamma_{hBN}z}-Fe^{+j\gamma_{hBN}z}\right) & s+d<z<s+d+t \\ \gamma_0\left(Ge^{-j\gamma_0 z}\right) & z>s+d+t \end{cases} \tag{10}$$

In (9)-(10), the coefficients $A, B, C, D, E, F, G$ are unknown coefficients and will be determined by applying the following boundary conditions:

$$E_{x,2}=E_{x,1}=E_x \tag{11}$$

$$H_{y,2}-H_{y,1}=\begin{cases} \sigma E_x & if\ z=s+d+t \\ 0 & otherwise \end{cases} \tag{12}$$

$$E_x=0\ \ @\ z=0\ \ \Rightarrow\ \ A=B \tag{13}$$

Finally, we obtain a matrix representation by applying the above boundary conditions,

$$\begin{pmatrix} a_1 & a_2 & a_3 & 0 & 0 & 0 \\ 0 & a_4 & a_5 & a_6 & a_7 & 0 \\ 0 & 0 & 0 & a_8 & a_9 & a_{10} \\ a_{11} & a_{12} & a_{13} & 0 & 0 & 0 \\ 0 & a_{14} & a_{15} & a_{16} & a_{17} & 0 \\ 0 & 0 & 0 & a_{18} & a_{19} & a_{20} \end{pmatrix} \cdot \begin{pmatrix} A \\ C \\ D \\ E \\ F \\ G \end{pmatrix} = \begin{pmatrix} 0 \\ 0 \\ 0 \\ 0 \\ 0 \\ 0 \end{pmatrix} \tag{14}$$

In (14), the following coefficients have been used:

$$a_1=-2j\frac{\gamma_{Si}}{\varepsilon_{Si}}\sin\left(\gamma_{Si}s\right),\ a_2=-\frac{\gamma_{SiO_2}}{\varepsilon_{SiO_2}}e^{-j\gamma_{SiO_2}s},\ a_3=\frac{\gamma_{SiO_2}}{\varepsilon_{SiO_2}}e^{+j\gamma_{SiO_2}s},\ a_4=\frac{\gamma_{SiO_2}}{\varepsilon_{SiO_2}}e^{-j\gamma_{SiO_2}(s+d)} \tag{15}$$

$$a_5=-\frac{\gamma_{SiO_2}}{\varepsilon_{SiO_2}}e^{+j\gamma_{SiO_2}(s+d)},\ a_6=-\frac{\gamma_{hBN}}{\varepsilon_{hBN}}e^{-j\gamma_{hBN}(s+d)},\ a_7=\frac{\gamma_{hBN}}{\varepsilon_{hBN}}e^{+j\gamma_{hBN}(s+d)},\ a_8=\frac{\gamma_{hBN}}{\varepsilon_{hBN}}e^{-j\gamma_{hBN}(s+d+t)} \tag{16}$$

$$a_9=-\frac{\gamma_{hBN}}{\varepsilon_{hBN}}e^{+j\gamma_{hBN}(s+d+t)},\ a_{10}=-\gamma_0 e^{-j\gamma_0(s+d+t)},\ a_{11}=2\cos\left(\gamma_{Si}s\right),\ a_{12}=-e^{-j\gamma_{SiO_2}s} \tag{17}$$

$$a_{13}=-e^{+j\gamma_{SiO_2}s},\ a_{14}=e^{-j\gamma_{SiO_2}(s+d)},\ a_{15}=e^{+j\gamma_{SiO_2}(s+d)},\ a_{16}=-e^{-j\gamma_{hBN}(s+d)} \tag{18}$$

$$a_{17} = -e^{+j\gamma_{hBN}(s+d)},\ a_{18} = e^{-j\gamma_{hBN}(s+d+t)},\ a_{19} = e^{+j\gamma_{hBN}(s+d+t)},\ a_{20} = -\left(1+\frac{\sigma\gamma_0}{\omega\varepsilon_0}\right)e^{-j\gamma_0(s+d+t)} \tag{19}$$

Setting $det(coeffcients) = 0$ will give us the dispersion relation:

$$\begin{aligned} &f_1(\omega).\sin(\gamma_{Si}s)\cos(\gamma_{hBN}t)\cos(\gamma_{SiO_2}d) + f_2(\omega).\sin(\gamma_{Si}s)\sin(\gamma_{hBN}t)\sin(\gamma_{SiO_2}d) + \\ &f_3(\omega).\sin(\gamma_{Si}s)\sin(\gamma_{hBN}t)\cos(\gamma_{SiO_2}d) + f_4(\omega).\sin(\gamma_{Si}s)\cos(\gamma_{hBN}t)\sin(\gamma_{SiO_2}d) - \\ &f_5(\omega).\cos(\gamma_{Si}s)\cos(\gamma_{hBN}t)\sin(\gamma_{SiO_2}d) + f_6(\omega).\cos(\gamma_{Si}s)\sin(\gamma_{hBN}t)\sin(\gamma_{SiO_2}d) - \\ &f_7(\omega).\cos(\gamma_{Si}s)\cos(\gamma_{SiO_2}d) - f_8(\omega).\cos(\gamma_{Si}s)\cos(\gamma_{SiO_2}d) = 0 \end{aligned} \tag{20}$$

In (20), the following coefficients have been used:

$$f_1(\omega) = -2j\frac{\gamma_{Si}}{\varepsilon_{Si}}\frac{\gamma_{hBN}}{\varepsilon_{hBN}}\frac{\gamma_{SiO_2}}{\varepsilon_{SiO_2}}\left(1+\frac{\sigma\gamma_0}{\omega\varepsilon_0}\right) \tag{21}$$

$$f_2(\omega) = 2j\frac{\gamma_{Si}}{\varepsilon_{Si}}\left(\frac{\gamma_{hBN}}{\varepsilon_{hBN}}\right)^2\left(1+\frac{\sigma\gamma_0}{\omega\varepsilon_0}\right) \tag{22}$$

$$f_3(\omega) = 2\gamma_0\frac{\gamma_{Si}}{\varepsilon_{Si}}\frac{\gamma_{SiO_2}}{\varepsilon_{SiO_2}} \tag{23}$$

$$f_4(\omega) = 2\gamma_0\frac{\gamma_{Si}}{\varepsilon_{Si}}\frac{\gamma_{hBN}}{\varepsilon_{hBN}} \tag{24}$$

$$f_5(\omega) = 2j\left(\frac{\gamma_{SiO_2}}{\varepsilon_{SiO_2}}\right)^2\frac{\gamma_{hBN}}{\varepsilon_{hBN}}\left(1+\frac{\sigma\gamma_0}{\omega\varepsilon_0}\right) \tag{25}$$

$$f_6(\omega) = 2\gamma_0\left(\frac{\gamma_{SiO_2}}{\varepsilon_{SiO_2}}\right)^2 \tag{26}$$

$$f_7(\omega) = \frac{\gamma_{SiO_2}}{\varepsilon_{SiO_2}}\frac{\gamma_{hBN}}{\varepsilon_{hBN}}\left(\gamma_0 + \frac{\gamma_{hBN}}{\varepsilon_{hBN}}\left(1+\frac{\sigma\gamma_0}{\omega\varepsilon_0}\right)\right)e^{-j2\gamma_{hBN}(s+d)} \tag{27}$$

$$f_8(\omega) = \frac{\gamma_{SiO_2}}{\varepsilon_{SiO_2}}\frac{\gamma_{hBN}}{\varepsilon_{hBN}}\left(\gamma_0 - \frac{\gamma_{hBN}}{\varepsilon_{hBN}}\left(1+\frac{\sigma\gamma_0}{\omega\varepsilon_0}\right)\right)e^{+j2\gamma_{hBN}(s+d)} \tag{28}$$

Now, our analytical model is completed for the proposed heterostructure and obtaining the propagation features such as the effective index ($N_{eff} = Re[\beta]/k_0$), the propagation length ($L_{Prop} = 1/Im[\beta]$) and the Figure of Merit ($FOM = Re[\beta]/Im[\beta]$) [89] is straightforward.

## 3. Results and Discussions

In this section, we will study the analytical results of the presented model in the previous section. To simulate the structure, the chemical potential of graphene is set to be $\mu_g = 0.25\ ev$, the temperature is $T = 300\ K$, the relaxation time is $\tau = 0.45\ ps$, and the thickness of graphene is supposed to be $\Delta = 1\ nm$. The parameters of the hBN medium are given in the previous section. The permittivity constant of $SiO_2$ and Si layers are 2.09 and 11.9, respectively. The geometrical parameters are assumed to be $t = 10nm, d = 20nm, s = 30\ nm$. The structure has been simulated in COMSOL by using the finite element method (FEM) and the simulation results will be compared with analytical ones to check the validity of the proposed model. The perfectly matched layers (PML) have been utilized as the boundary conditions in our simulation. The mesh sizes of the structure are supposed to be $\Delta x = \Delta y = \Delta z = 0.2$ nm in all directions.

As explained before, the hBN medium shows two various phonon modes: out-of-plane (type-I) and in-plane (type-II) phonon modes. In our study, the interest frequency region is type-II hyperbolicity because of $Re[\varepsilon_{\parallel}] < 0$ in this region (similar to the negative permittivity of graphene sheet in the mid-infrared frequencies). Thus, we focus on the upper Reststrahlen band. In fig. 3, we have illustrated the simulation and analytical results of the effective index and propagation length as a function of frequency. As seen in this figure, there is a good agreement between the simulation and analytical results, which confirms the validity of the mathematical expressions. The effective index increases as the frequency increases while there is an optimum value for the propagation length which occurs around 43.5 THz.

To effectively study the performance of the proposed heterostructure, we will investigate the figure of merit (defined by $FOM = Re[\beta]/Im[\beta]$ [89]) of the structure here. This parameter will help us to better understand the performance of $SP^3$ inside the structure. Fig. 4 represents the FOM variations as a function of frequency for various values of hBN thickness and chemical potential. Other parameters have remained fixed. It can be observed from fig. 4(a) that as the thickness of the hBN medium increases, the FOM decreases because the confinement of $SP^3$ reduces. Moreover, FOM increases as the frequency increases for the frequency range of $\omega < 45\,THz$, while it begins to decrease with the frequency increment for $\omega > 45\,THz$. One can see from fig. 4(b) that the increment of the chemical potential increases FOM because of the increment of the propagation length for the higher values of chemical potential. Furthermore, the slope of variations increases for the higher values of chemical potential. A high value of FOM=190 is observable for $\mu_c = 0.85\,ev$ at the frequency of 48.3 THz. Compared to the previous articles [90-93], our achieved FOM shows a remarkable increment. In [90], a fabricated graphene-based waveguide is investigated with a reported FOM of 33. In [91], a quality factor of 25 is obtained for a hybrid graphene-hBN heterostructure and the authors in [92] have reported FOM=45 for their graphene-based nano-wire.

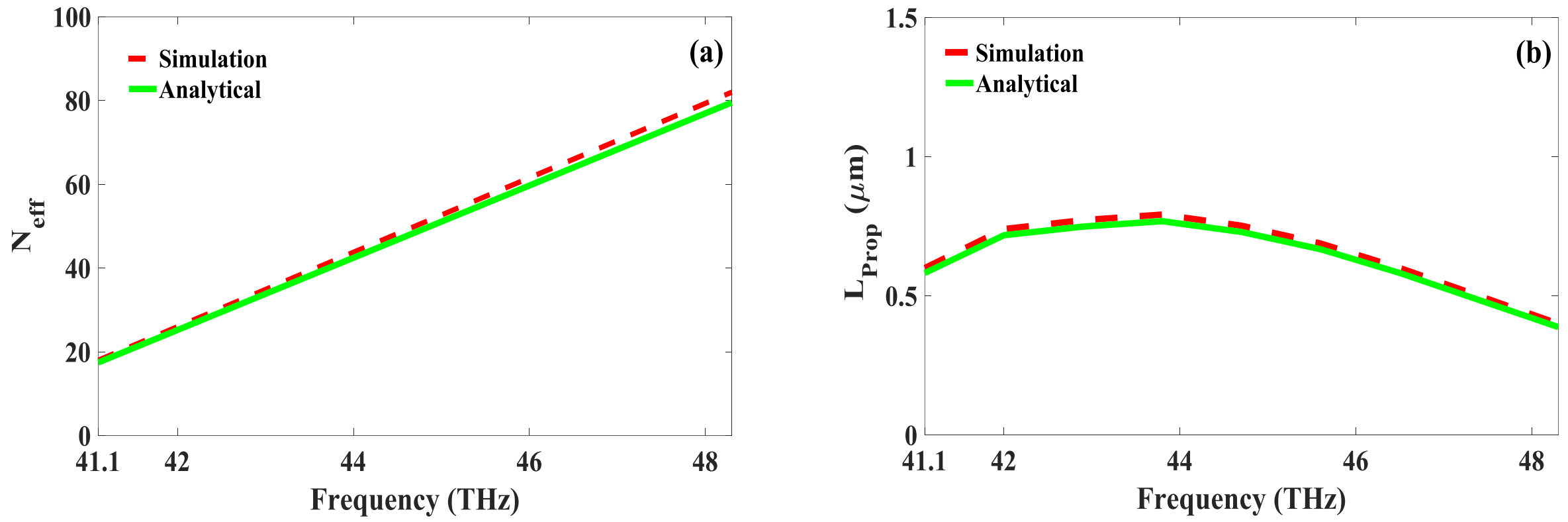

**Fig. 3.** Propagating properties of the studied structure versus frequency: **(a)** the effective index**, (b)** the propagation length. The solid and dashed lines show the analytical and simulation results, respectively. The chemical potential is supposed to be 0.25 eV. The thickness of the hBN layer is 10 nm.

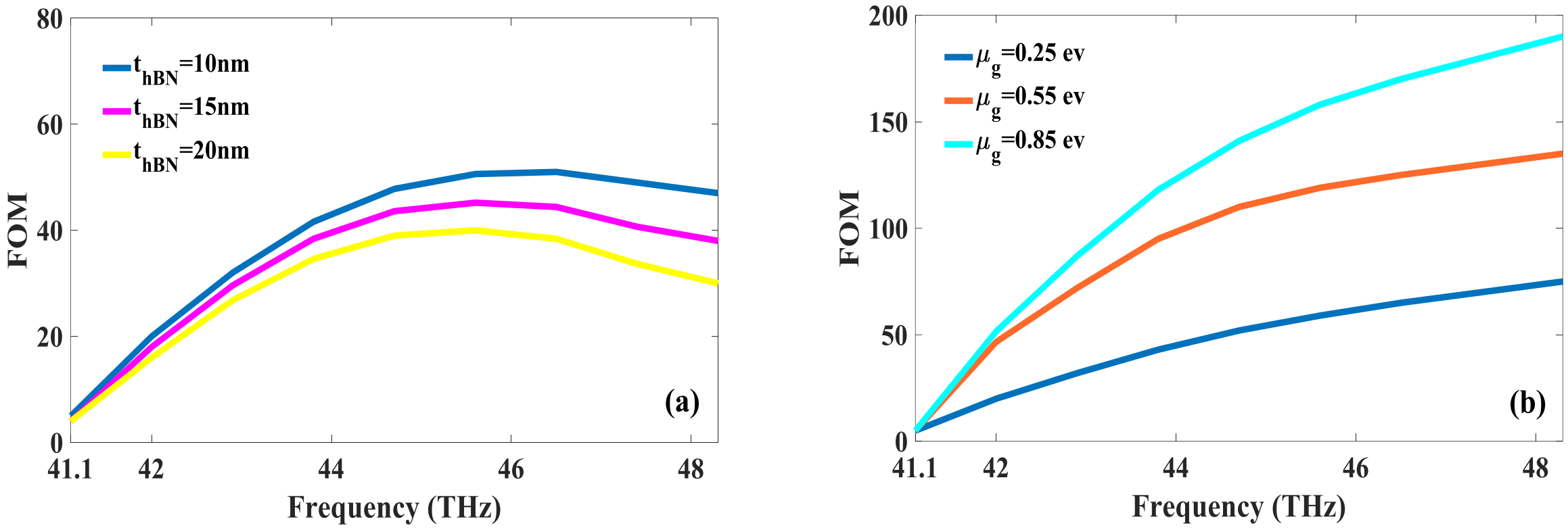

**Fig. 4.** The dependence of FOM on frequency for various values of: **(a)** the hBN thickness, **(b)** the chemical potential of graphene. In fig. 4(a), the chemical potential is 0.25 eV and the thickness of the hBN layer is assumed to be 10 nm in fig. 4(b).

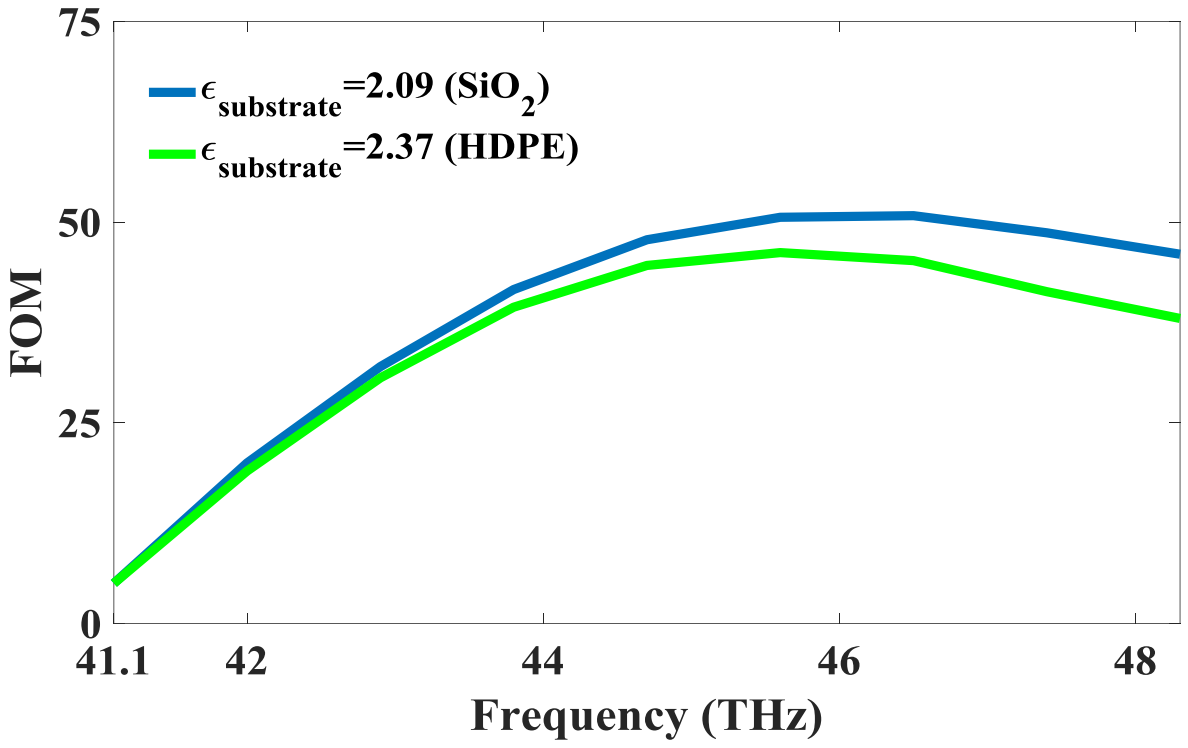

**Fig. 5.** FOM versus frequency for various values of substrate permittivity (mid-layer). The chemical potential is supposed to be 0.25 eV. The thickness of the hBN layer is 10 nm.

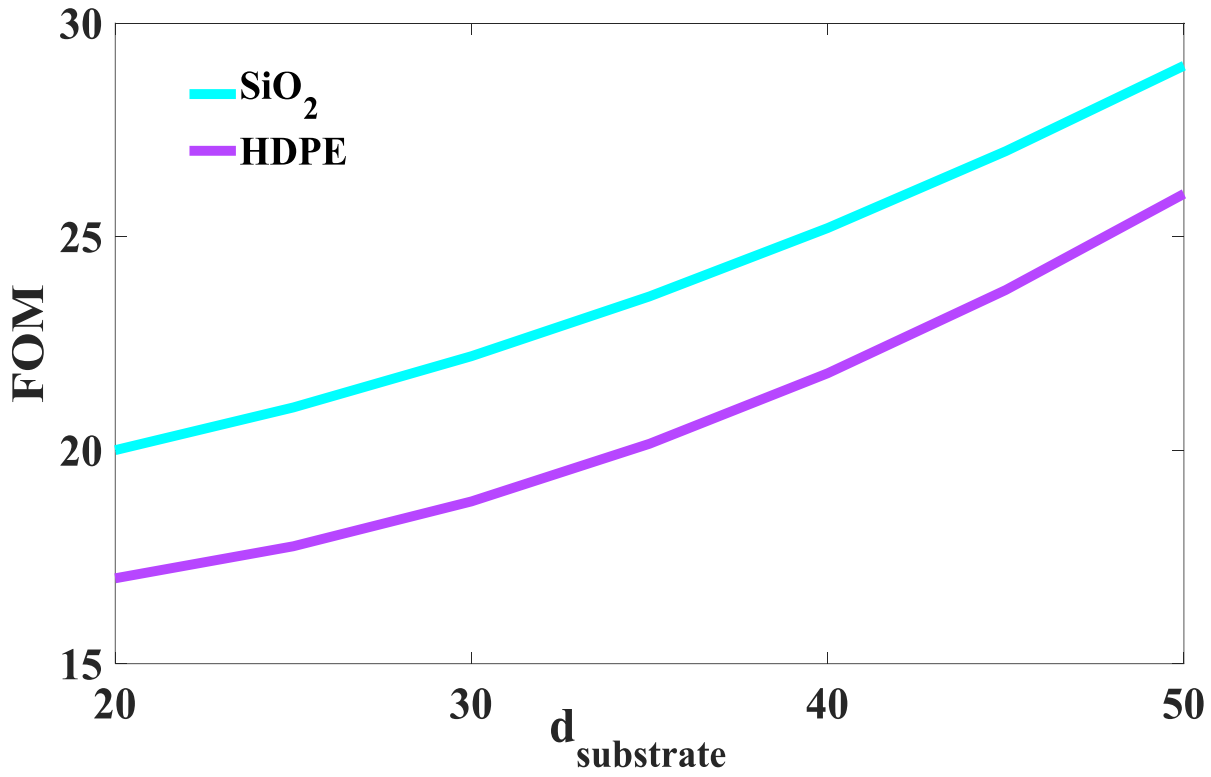

**Fig. 6.** The dependence of FOM on the thickness of the substrate (d). The chemical potential is supposed to be 0.25 eV. The thickness of the hBN layer is 10 nm. The frequency is 42 THz.

One of the important layers in our proposed device in which its parameters can change the performance is the middle substrate, which is primarily supposed to be a $SiO_2$ layer with a thickness of $d = 20nm$. However, it is worth changing the characteristics of the middle substrate (such as permittivity and thickness) to enhance the quality of propagating SP[3]. Here, we report and compare the analytical results for two substrates: $SiO_2$ and High-Density Polyethylene (HDPE). It should be noted that the previous mathematical expressions are also valid for a heterostructure with an HDPE substrate by only substituting $\varepsilon_{SiO_2} \rightarrow \varepsilon_{HDPE}$. As seen in fig. 5, FOM decreases as the permittivity of the substrate increases. This happens because as the permittivity increases, the confinement and the propagation loss increase (the propagation length decreases), which finally will increase the FOM of the structure.

Fig. 6 shows the FOM variations as a function of substrate thickness for various kinds of the middle substrate ($SiO_2$, HDPE). In this diagram, the thickness varies from 20 to 50nm. As the thickness of the middle substrate increases, FOM enhances slightly. Moreover, the usage of a substrate with a higher value of permittivity will decrease FOM.

To further investigate the influence of chemical potential on performance, the FOM has been illustrated in fig. 7 as a function of chemical potential when it varies from 0.25 to 1 eV. In this figure, it is supposed that the middle substrate is $SiO_2$ with a thickness of 20 nm. The studied frequency is 42 THz. It can be found from fig. 7 that FOM increases as the chemical potential increases. Moreover, higher values of hBN thickness result in lower FOM. For instance, FOM=71 is achievable for the chemical potential of 0.95 eV at the frequency of 42 THz.

As a final point, the influence of the thickness of the bottom substrate (i.e. $S_{Si}$) on FOM has been demonstrated in fig. 8. In this figure, the thickness of the Si layer varies from 30 to 60 nm. The thickness of the middle substrate is 20 nm (the middle layer is supposed to be $SiO_2$). The studied frequency is 42 THz here. It can be observed that the slope of FOM variations is slight, which means that the change of $S_{Si}$ has a negligible effect on the performance of the structure. For instance, consider the chemical potential of 0.25 eV. As $S_{Si}$ varies from 30 to 60 nm, the FOM changes from 20 to 25; thus the performance of the proposed structure is not very sensitive to the variations of $S_{Si}$. Moreover, one can see from fig. 8 that higher values of chemical potential will increase FOM.

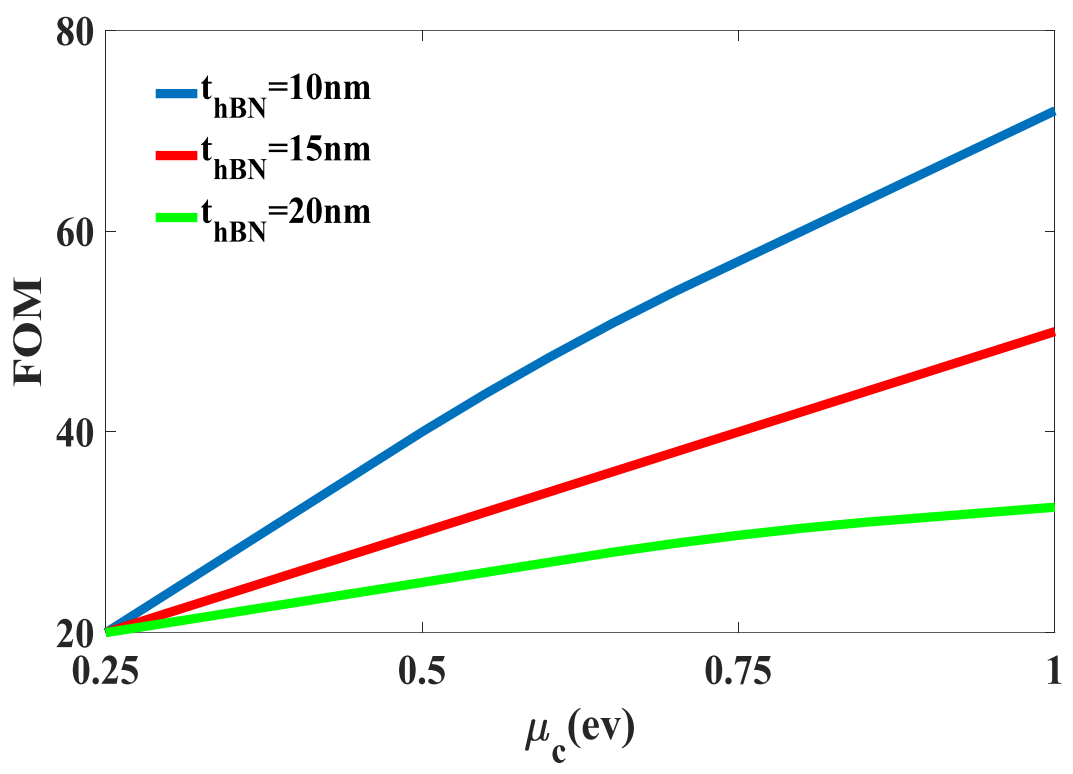

**Fig. 7.** The dependence of FOM on the chemical potential of graphene for various values of hBN thickness. The thickness of the substrate is 20 nm (the middle layer is supposed to be $SiO_2$). The frequency is 42 THz.

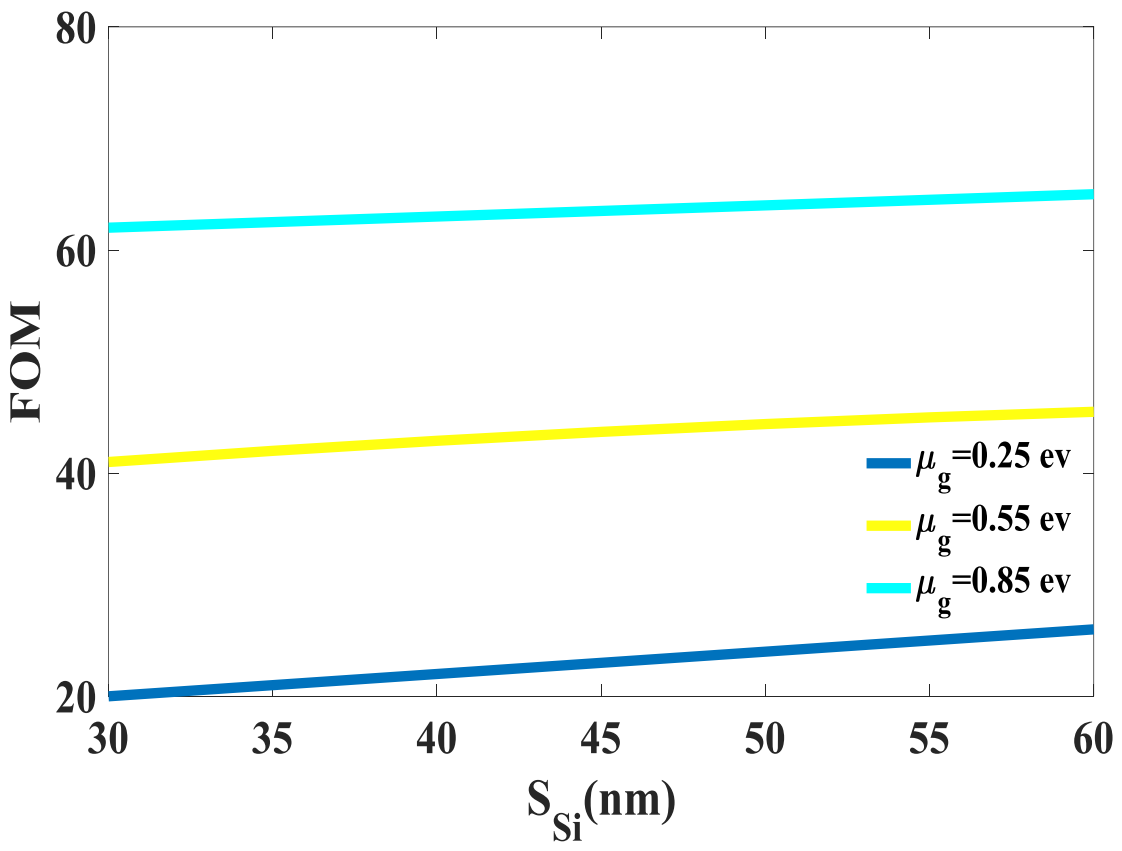

**Fig. 8.** The dependence of FOM on the thickness of the Si layer for various values of chemical potential. The thickness of the substrate is 20 nm (the middle layer is supposed to be $SiO_2$). The frequency is 42 THz. The thickness of the hBN layer is 10 nm.

## 4. Conclusion

In this paper, an analytical model was proposed for a new graphene-based hBN heterostructure supporting tunable $SP^3$. The model was started with Maxwell's equations and then applied boundary conditions. An exact dispersion relation was derived for the proposed structure in which the comparison between simulation and analytical results confirmed its validity. A high value of FOM=190 was reported for the chemical potential of 0.85 eV at the frequency of 48.3 THz. To further show the tunability of the structure, the influence of chemical potential and other geometrical parameters on the quality of propagating $SP^3$ were investigated in detail. We believe that the presented study can be useful for the design of novel graphene-based devices in the THz region.

## Declarations

**Ethics Approval:** Not Applicable.

**Consent to Participate:** Not Applicable.

**Consent for Publication:** Not Applicable.

**Funding:** The authors received no specific funding for this work.

**Conflicts of Interest/ Competing Interests:** The authors declare no competing interests.

**Availability of Data and Materials:** Not Applicable.

**Code availability:** Not Applicable.

**Authors' Contributions:** M. B. Heydari proposed the main idea of this work and performed the analytical modeling. M. Karimipour conducted the numerical simulations and wrote the manuscript. M. Mohammadi Shirkolaei analyzed the results and reviewed the paper.